# The Design of EzWindows:
# A Graphics API for an Introductory Programming Course

Bruce R. Childers, James P. Cohoon, Jack W. Davidson[†], Peter Valle

*Department of Computer Science*
*University of Virginia*
*Charlottesville, Virginia 22903*

{brc2m,cohoon,jwd,prv4d}@cs.virginia.edu

## Abstract

*Teaching object-oriented programming in an introductory programming course poses considerable challenges to the instructor. An often advocated approach to meeting this challenge is the use of a simple, object-oriented graphics library. We have developed a simple, portable graphics library for teaching object-oriented programming using* `C++`*. The library, EzWindows, allows beginning programmers to design and write programs that use the graphical display found on all modern desktop computers. In addition to providing simple graphical objects such as windows, geometric shapes, and bitmaps, EzWindows provides facilities for introducing event-based programming using the mouse and timers. EzWindows has proven to be extremely popular; it is currently in use at over 200 universities, colleges, and high schools. This paper describes the rationale for EzWindows and its high-level design.*

## 1 Introduction

Teaching object-oriented programming in an introductory programming course poses considerable challenges to the instructor [3]. In addition to covering the basic programming constructs (e.g., naming, fundamental types, expression evaluation, assignment, input/output, program control, functions, etc.), the instructor also must introduce the elementary mechanics of O–O programming such as object creation, object manipulation, and object interaction as well as simple object-oriented design concepts and approaches. Integrating these concepts in an introductory course is challenging. An often advocated approach to meeting this challenge is the use of a simple, O–O graphics library [1]. Despite the obvious appeal and advantages of using graphics in an first programming course, few such libraries are available for instructors to use. This paper describes the rationale for and design of a simple portable, `C++` graphics API (application programmer interface) called EzWindows. In addition to providing basic graphical objects such as windows and geometric shapes, the EzWindows API provides simple interfaces to the mouse and the system clock. The mouse interface allows beginning programmers to use the mouse as an input device. The interface to the system clock allows beginning programmers to develop programs that respond to timer events. Together, the mouse and clock interfaces provide a natural basis for introducing event-based programming which is a perfect framework for demonstrating the power of O–O design and implementation.

EzWindows has proven to be extremely popular; it is in use at over 200 universities, colleges, and high schools. Several factors have contributed to EzWindows' popularity. EzWindows is simple, yet it allows the student to exploit the input and output capabilities of desktop computers. The ability to create graphical objects and have them respond to messages programmed by the students really does seem to help teach O–O programming concepts. Using an API such as EzWindows introduces students to the real-world practice of developing programs using application-specific libraries. Finally, we designed EzWindows to be platform and compiler independent. Versions of EzWindows are available for:

- Windows 95/98/NT with all the major `C++` compilers (e.g, Borland `C++`, Microsoft Visual `C++`, Metrowerks CodeWarrior, and IBM VisualAge `C++`),
- Unix/Linux and X11 for many platforms using the vendor-supplied `C++` compiler and GNU's `g++`, and
- Macintosh/PowerPC using Metrowerks `C++` CodeWarrior.

---
[†]Corresponding author.

## 2 EzWindows design

### 2.1 EzWindows rationale

Initially our single objective was to develop a simple graphics library that would permit beginning programmers to write programs that used the graphical display found on modern desktop computers. This objective had two primary motivations.

First, our early experience with teaching O–O design and implementation in our introductory programming course (our department began using `C++` in 1991) was that students have a difficult time understanding the object concept when the only objects that they can manipulate are abstractions of objects: cars, bank accounts, dice, fractions, employees, etc. To help the students better grasp the O–O paradigm, we wanted to provide the students the ability to create concrete objects that they could manipulate and see the results of their manipulations.

Second, to better motivate today's students, we strongly believed that we needed to offer students the ability to write programs that did more than just read data, compute something, and print the result. Our experience with teaching students who had already been exposed using a desktop computer with a modern user interface was that many students quickly became bored and frustrated because they could not write programs that operated in a similar manner. Frankly, we were somewhat bored too.

As we explored developing a library for supporting simple, graphical programming, we realized that this was an opportunity to make significant changes in what we taught in our introductory courses. If we were going to provide facilities for the students to develop programs that produced graphical output, it seemed natural to also provide support for input via the mouse. Initially, we were somewhat wary that we were going too far, and that it would be difficult to package and implement the needed capabilities in a way that would be easily accessible to students. We were also concerned about portability to other platforms and differences in the various compilers available. We did not want to lock ourselves in to one particular platform or compiler. However, as we investigated the internal workings of Microsoft Windows, we became convinced that with careful design and implementation that it would be possible to provide a package that would provide simple graphical output and input via the mouse.

If we could use the mouse for input, this afforded us the opportunity to introduce event-based programming to our students. If used judiciously, event-based programming could help the students grasp the power of O–O design. Also the handling of events is the underlying paradigm in most user interfaces and many simulations. This seemed like an innovative and exciting possibility and we tentatively decided to pursue introducing event-based programming in our beginning programming course.

Finally, we hope to build EzWindows so that itself could be used as an example of good O–O design and a source of examples illustrating O–O implementation using `C++`.

### 2.2 EzWindows architecture

Now that we had the general vision of the capabilities we wanted in EzWindows, we had to determine the overall architecture and the detailed requirements. As practicing programmers ourselves, we used application-specific libraries to develop software. We observed that programmers rarely build a software application from scratch. Most real-world software is built using application-specific libraries or APIs. APIs exist for a variety of applications: GUI development, 3-D graphics, virtual reality, telephony, database access, spreadsheet access, cryptography, network programming, etc. Indeed, a common programming activity is learning what capabilities an API provides and how to use it. Based on these observations, we decided to structure and present EzWindows as an API. Essentially the students would be presented with a set of objects and their public interfaces. Their implementations would be hidden initially. Structuring EzWindows as a set of objects had the added benefit that if we did a good job of designing the EzWindows object interfaces, they would serve as a model of good object design. We strongly believe that beginning students should be client users of objects before they can become competent designers.

Another early decision we had to make was whether to use any of the existing libraries provided by the compiler vendors to implement EzWindows. We had been using Borland `C++` compiler in our introductory courses so we had some experience with its Object Works Library (OWL). Another possibility was to use Microsoft Founda-tion Classes (MFC) as it was gaining wide acceptance. On the plus side, if we used OWL or MFC the implementation of EzWindows would be simpler. On the negative side, it would have made it difficult to make EzWindows both compiler and platform independent. Ultimately we decided that it would be acceptable if EzWindows was limited to running on Wintel machines, but compiler independence was mandatory. We were currently a Borland shop, but we knew it was possible that circumstances beyond our control might require us to use a different compiler.

The decision not to use a library supplied by a compiler vendor meant that EzWindows would have to be implemented on top of the basic Windows API. The Windows API is a set of low-level services (e.g., process creation, window creation, graphics primitives, etc.) that are supplied by all Windows development environments. The low-level nature of the Windows API would complicate the implementation of EzWindows; however, it would also mean that any services we used could, with modest effort, be implemented on other platforms such as Unix/X11 or the Macintosh. Fortunately, this turned out to be true. Having settled on the overall architecture and the underlying library we would use for implementation, we were now ready to begin the design of the EzWindows classes.

### 2.3 EzWindows classes

To insure that EzWindows provided a simple vehicle for teaching beginning programmers O–O programming fundamentals, we decided to design the high-level classes that the students would initially use first. We would focus on keeping the interfaces and the implementation of the classes as simple as possible. The resulting design of these classes would then drive the design and implementation of the low-level portion of EzWindows. Figure 1 shows the implementation structure of EzWindows.

Using this layered approach had two important benefits. First, it would give us greater control over the implementation of the high-level classes. We did not want the implementation of the high-level EzWindows classes to be so complicated that our students would not be able to understand it. Second, by imposing a middle layer, the implementation of the high-level classes would be platform independent. Porting EzWindows to a new platform would only require providing a different implementation of the low-level EzWindows classes and primitives. The client view (i.e., the student view) would be identical across all platforms.

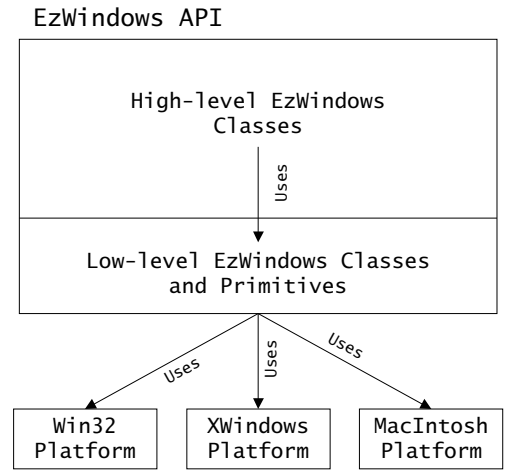

**Figure 1:** EzWindows implementation structure.

The central object in EzWindows is `SimpleWindow`. Because all EzWindows' graphical objects would be bound to and displayed in a `SimpleWindow`, it would be one of the first objects students would see and use. Consequently, the `SimpleWindow` class should be as simple possible. Two of the key design issues for `SimpleWindow` were the coordinate systems to use for positioning a `SimpleWindow` on the screen and objects within a `SimpleWindow`, and the unit of measure to use for specifying the size of a `SimpleWindow` and other objects.

Most windowing systems use a coordinate system based on pixels for positioning a window on the screen. Typically the origin is the upper-left corner of the screen with the X-axis extending horizontally to the right and the Y-axis extending vertically down. A window position is specified by giving the coordinates of its upper-left corner.

There are several problems with using pixels as the basis of a simple coordinate system. First, the observed position depends on the resolution of the display device; a window at (20, 30) on a system with a 800 by 600 pixel display would be rendered farther from the upper-left corner of the screen than on a system with a 1600 by 1200 pixel display. Second, the aspect ratio of a computer display is not 1:1. The aspect ratio is resolution dependent. Thus, a window positioned at (400, 400) would not be rendered the same distance from the origin along the X- and the Y-axis.

To avoid these complications, we decided to use distance in centimeters from the origin along the X- and Y-axis as the coordinate system. Thus, the coordinate (3.0, 4.0) specifies a location 3 centimeters from the left edge of the screen and 4 centimeters from the top of the screen. We also used centimeters to specify the size of a `SimpleWindow` and other objects. We have found that using this coordinate system and method for specifying the sizes of objects helps beginning programmers produce displays that correspond to their mental model more quickly than a system based on pixels. Figure 2 shows the EzWindows coordinate system.

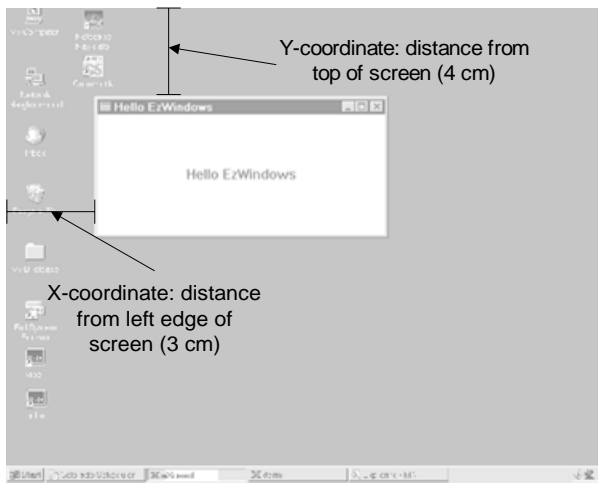

**Figure 2:** EzWindows coordinate system.

Two types of objects can be displayed within a `SimpleWindow`: 2-dimensional shapes and bitmaps. The shape classes provided by EzWindows include `RectangleShape`, `EllipseShape`, and `TriangleShape`. Many graphics system use a bounding box for specifying the size and position of a shape in the containing window. The size of the shape is specified by giving the width and height of a bounding box, and the position of the shape is specified by giving the screen coordinate of the upper-left corner of the bounding box. Figure 3 shows the bounding box for specifying an ellipse.

We debated whether to use the above approach for specifying the position of a shape, or to use the center of the shape as the positioning point. We flip flopped on this issue several times. Ultimately, we chose to use the center of the shape as the positioning point because we believed the students would find it easier to understand initially. Interestingly, when we added bitmaps to the EzWindows library we decided to use the upper-left corner of the bounding box for the positioning point. This design inconsistency has bothered some users (and justifiable so). In the course we teach, we use this inconsistency as an example of poor design[2].

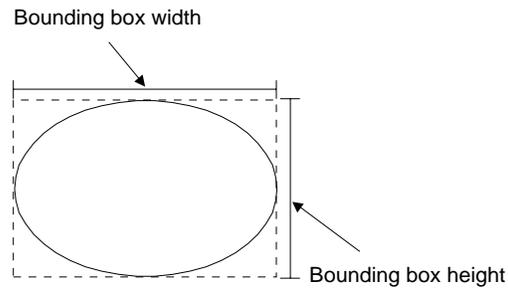

**Figure 3:** Bounding box for an ellipse.

One of our goals with EzWindows was to use its design and implementation as a source of examples for our students throughout the course. The advantage is that they would already familiar with using EzWindows, thus they can focus on its design and implementation. For example, the EzWindows shapes hierarchy has proven ideal for explaining and demonstrating the utility of inheritance (see Figure 4). Similarly, the implementation of the EzWindows shapes is a rich source of concrete, relevant examples of important principles such as encapsulation, modularity, and abstraction. It is also a source of examples for demonstrating how the features of `C++` can be used to build flexible interfaces.

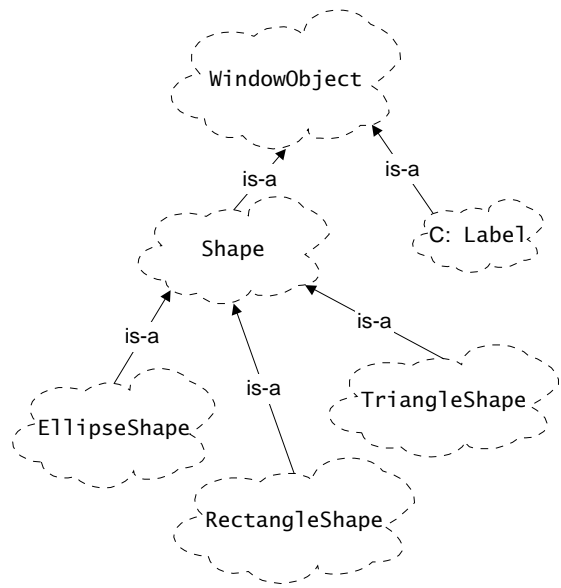

**Figure 4:** EzWindows shapes hierarchy.

For example, the power of information hiding is concretely illustrated by reviewing the low-level primitives and showing how the mechanisms for positioning and sizing objects are hidden from the client. The utility of function overloading and default parameters is concretely illustrated by showing how interfaces can be refined and expanded to include new features without requiring previously written client code to be modified.

## 3 EzWindows event architecture

EzWindows is unique in that it provides support so students can build programs that respond to external events generated via the mouse or by user-defined timers. This support allows the student to write programs that accept input using the mouse and to write programs that evolve over time (e.g., simple animations, simulations, etc.). The EzWindows event model and handling is very simple (see Figure 5).

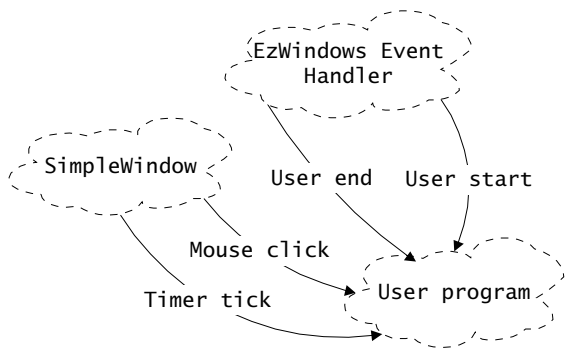

**Figure 5:** EzWindows event handling.

Programs that use EzWindows receive a `UserStart` message when the program begins execution and a `UserEnd` message immediately before the program is terminated. The `UserStart` message is sent to a function called `ApiMain`. For simple programs that only use the display, `UserStart` can be the only message that the client program handles (a default handler is supplied for `UserEnd`). Class `SimpleWindow` contains methods for registering mouse event and timer event callbacks. For example, assuming a `SimpleWindow` object name `W1` has been instantiated, the method invocation

```
W1.SetMouseClickCallback(MouseEvent);
```

will cause the EzWindows system to call function `MouseEvent` when the mouse is clicked within `SimpleWindow W1`. The function `MouseEvent` is passed the coordinates of the mouse sprite within the window. Timer events are registered similarly, however, there are functions for setting the interrupt interval and starting and stopping the timer.

With modest effort, this simple system of events allows students to build programs with polished graphical interfaces. Indeed, we have been surprised at the quality of the programs students have built.

## 4 Summary

This paper has described the rationale and design of EzWindows—a graphics API for a beginning object-oriented programming course. EzWindows has proven to be extremely popular. Its popularity is due to several factors. While simple to use, EzWindows allows beginning students to build surprisingly polished programs. This power helps motivate the students. Its O–O implementation serves as an excellent case study for students who had used it. It is platform and compiler independent. Finally, to the best of our knowledge it is the first graphics package that allows beginning programmers to use the mouse as an input device that they control.

## Acknowledgments

We thank the many users of EzWindows who have sent us comments and suggestions. Several users deserve special thanks: Bill Slough and Peter Andrews of Eastern Illinois University made many suggestions which were incorporated in EzWindows 2.0. We also thank Devon Lockwood for implementing the UNIX/X11 version of EzWindows. This work has been supported in part by the NSF under grants CDA–9634333, NSF–CS-5239-92, and DUE–9554715.